\documentclass[review]{elsarticle}

\usepackage{color}
\usepackage{graphicx}
\usepackage{amsmath, amsthm, amssymb}
\usepackage{tikz}
\usepackage{pgf,tikz-cd}
\usepackage{pstricks-add}
\usetikzlibrary[patterns]
\usetikzlibrary{arrows}
\usetikzlibrary{calc,backgrounds}
\usepackage{caption}
\usepackage{pgfplots}
\usepackage{mathrsfs}
\usepackage{algorithm2e}
\newtheorem{theorem}{Theorem}[section]
\newtheorem{lemma}[theorem]{Lemma}
\newtheorem{definition}[theorem]{Definition}

\newtheorem{notation}[theorem]{Notation}

\newcommand{\nc}{\newcommand}
\newcommand\tta{\tilde\theta}
\nc{\on}{\mathrm{on}}%

\begin{document}

\title{Online exploration outside a convex obstacle}
\author[mymainaddress1]{Shai Gul }
\author[mymainaddress2]{Eitan Tiktinsky}
\author[mymainaddress2]{Slava Shamshanov}
\author[mymainaddress2]{Reuven Cohen}
\ead{reuven@math.biu.ac.il }
\address[mymainaddress1]{Department of Applied Mathematics, Holon Institute of Technology, Holon, 5810201, Israel}
\address[mymainaddress2]{Department of Mathematics, Bar-Ilan University, Ramat-Gan, 52900, Israel}
\begin{abstract}
A watchman path is a path such that a direct line of sight exists between each point in some region and some point along the path. Here, we study the online watchman path problem outside a convex polygon, i.e., in $\mathbb{R}^2\setminus \Omega$, where $\Omega$ is a convex polygon that is not known in advance.
We present an algorithm for the exploration of the region outside the polygon.
We prove that the presented algorithms guarantees a $\approx 22.77$ competitive ratio compared to the optimal offline watchman path.
\end{abstract}
\maketitle


\section{Introduction}
Exploring an unknown terrain or scanning a region of space are important tasks for autonomous robots. Many situations require exploring an unknown environment, or scanning a known environment for changes or intrusions. Some autonomous units, such as the mars rover \cite{ROVER,DBLP:conf/focs/DengKP91} and other space exploration vehicles are too far to control from earth, as the communication time is too long. In other cases communication is impossible due to interference or environmental conditions.

\begin{figure}[!htb]
\hspace*{-4cm}
    \centering
    \includegraphics[trim={2cm 0 8cm 0},clip,scale=0.6]{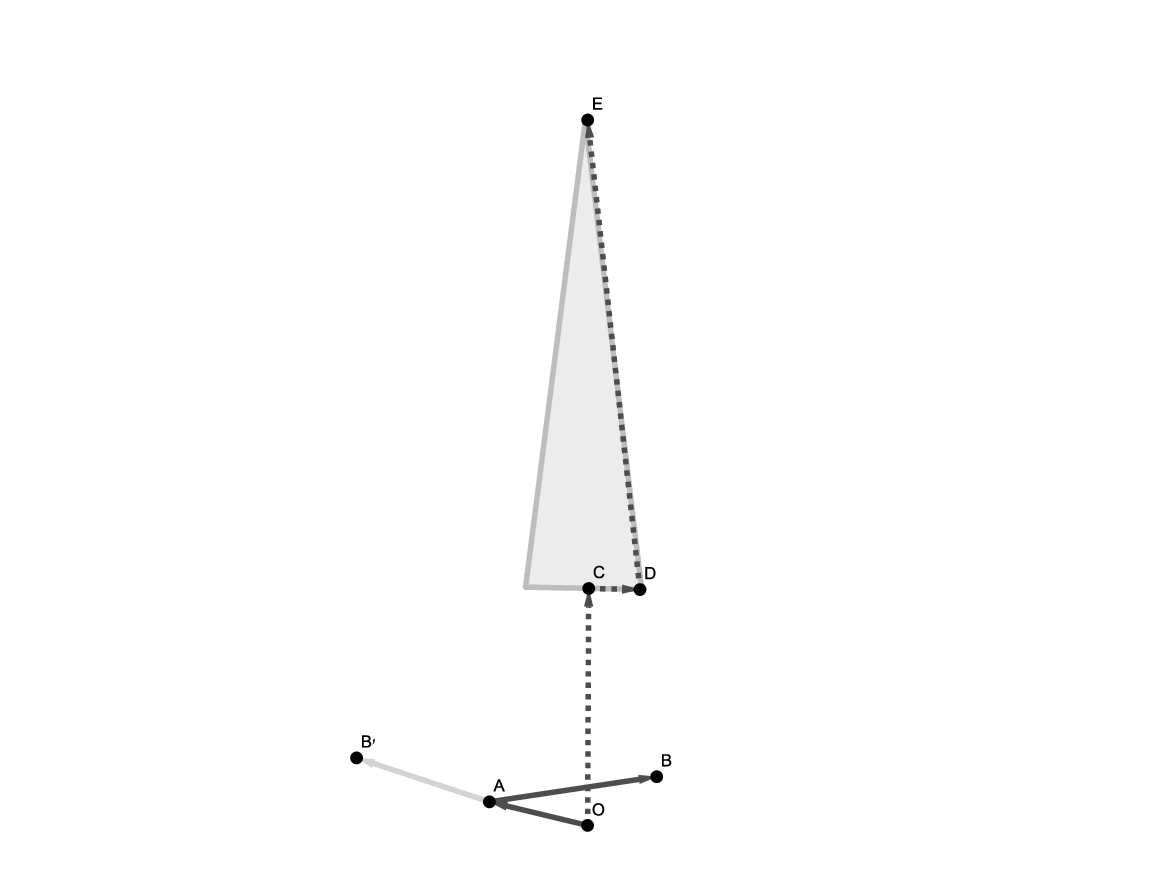}

\caption{The scanning starts from a point $O$. We illustrates two paths: the optimal path (for this specific case) is scanning along the edge $OA$ and then along the edge $AB$ where $AB$ is the reflection of $AB'$ along the line $EA$ and $AB$ is perpendicular to $EB$. The dotted path shows an intuitive path where we go toward the object and scan along its perimeter, which leads in narrow domain to a considerably longer path.}
\label{fig 2}
\end{figure}

In this paper, we study the problem of scanning or exploring a region of the plane, where a convex polygonal obstacle is blocking the view and motion of the robot (see Fig. (\ref{fig 2}). 
We present an online algorithms guaranteeing a constant competitive ratio compared to the optimal offline path. In the offline setting the shape of the obstacle is known in advance, and an optimal path is desired, where this optimal path is the shortest path from which every point in the free space (space outside the obstacle) is viewable (a direct line of sight exists).
Our main result is an algorithm for the online problem. In this case, the shape of the obstacle is not known in advance, and the purpose of the robot is to scan the region while studying the shape of the obstacle, where the goal is to minimize the length of the motion path. The algorithm we present guarantees a constant factor stretch of the motion path length, relative to the optimal solution of the offline problem.

 \section{Related work}

{\textit{The watchman's path problem}} \cite{CMWR,OGC,WRLV,ASWR,MP,Guibas1987} is a well-known optimization problem, where an algorithm for a watchman needs to be constructed so that he computes the shortest path to traverse a certain area, and from this path he must observe the entire area, we assume he has $2\cdot\pi$ view to any distance only bounded by obstacles. In the general case it has been shown to be an NP-hard problem \cite{MEGIDDO1982194,ARKIN2003203,OETO} (say if there are $n$ obstacles). In an offline scenario  the guard is given a map of the area including its obstacles and needs to compute the shortest path. Whereas, in the online case, the watchman has to explore (i.e. discover unknown terrains in the offline case) or scan (i.e. canvas a known terrain in the online case) the area it traverses without (or with limited) prior knowledge of what lays ahead .
		
    The watchman problem has been observed under various different constraints.
Online algorithms for touring  the interior of a non-convex simple polygon have been presented in \cite{OLSISP,PEP}. The traversing robot does not have a map, and is only aware of what it has explored so far. They present a  $\frac{5}{4}$-competitive algorithm. This problem is similar to our online setting. However, in their case, the polygon is simply connected, whereas in our case, the watchman tours the outside of a polygon, which is not simply connected.

Czyzowicz et al  \cite{OETO} give an algorithm for many obstacles (again, the obstacles may not be convex). However, even in a convex scenario, no constant competitive ratio is obtained.  In the exploration algorithm with unlimited vision, the complexity of the path is $O(P+D\cdot\sqrt{k})$ where $P$ is the total perimeter of the terrain (including obstacle perimeter), $D$ is the diameter of the convex hull of the terrain and $k$ is the number of obstacles. In \cite{DBLP:conf/focs/DengKP91} an algorithm for touring a general polygon is discussed,  a competitive ratio for touring the interior of a polygon is shown to be less than 2016, and the existence for a an online competitive algorithm for touring the exterior of a general polygon is also discussed, with no details on the ratio.

Georges	et al \cite{Georges2013} give an algorithm for touring a polygon with holes, where the holes are polygonal and are of different colors. Their algorithm is $\approx 600$-competitive in the case of one hole. Here, we give an improved competitive ratio for a convex obstacle (hole) in $\mathbb{R}
^2$.

\section{Foundations Layout}
\begin{definition}
A spiral logarithmic curve is a an infinite curve, defined by the polar coordinates
$$
r(\tta)=  e^{b \tta} \ ,
$$
where $\tta \in (-\infty, \infty)$ and $b$ is a constant (see Fig. \ref{fig:1}).
\end{definition}
The point $(\tta,r(\tta))$ corresponds to the Euclidean coordinates $x=r\cos\tta$, $y=r\sin\tta$. As the point is defined up to an
addition of an integer multiple $2\pi$ to $\tta$, we will use the notation $\theta=(\tta\mod 2\pi)$, where $\theta\in[0,2\pi)$.

\begin{notation}
Denote the length of a spiral from the origin $O$ to a distance $R$ (and an angle $\theta$) by $L_s(R)$.
\end{notation}
By directly calculating the integral, one obtains that the length $L_s(R)$ of the spiral up to a radius $R$,  is
\begin{align}
\label{length}
 L_s(R)= \sqrt{1+\frac{1}{b^2}}\cdot R
\end{align}
\begin{theorem}
\cite{gal_book} Let $x=(\theta,r)$ be a point and let $(\tilde{\theta},R)$ be the first point along the spiral such that $x$ lies on the segment between the origin and $(\tta,R)$ (i.e, ${R}\geq r$ and $\theta=\tta \mod 2\pi$). Then
\begin{align}
 L_s(R) \leq e^{2\pi b}\sqrt{1+\frac{1}{b^2}}r \ .
 \label{spiral}
\end{align}
\end{theorem}
A minimization of the expression (\ref{spiral}) leads to the competitive ratio
$L_s \leq 17.2894\cdot r$ between the direct line to a point and the length of the spiral up to the location on the infinite part of the ray from the origin to the point.

%


\begin{lemma}[\cite{EGC} p. $42$, Theorems $7.11$ and $7.12$.]
\label{boundary length}
Given two convex domains $\Omega_1$ and $\Omega_2$ with a respective perimeter $|\partial\Omega_1|$ and $|\partial\Omega_2|$.
If  $\Omega_1 \subseteq \Omega_2$, then $|\partial \Omega_1| \leq | \partial \Omega_2|$.

\end{lemma}

\section{Results}
We begin by showing that no online algorithm can obtain an approximation ratio better than $3$.
\begin{notation}
Denote the length of the optimal path by $L_\mathrm{opt}$ and the length of the online algorithm by $L_\mathrm{on}$
\end{notation}
\begin{theorem}
    For every $\varepsilon>0$, every online algorithm has a triangular obstacle for which its competitive ratio is at least $$\frac{L_{\on}}{L_\mathrm{opt}}  >3-\varepsilon \,.$$
\end{theorem}
\begin{proof}
    For any given triangle the adversary may place the robot at any starting point $O$.
    A bad possible option for the best possible online algorithm occurs when the triangle has two fairly long edges $e_{\ell_1}$ and $e_{\ell_2}$ with length $\ell$ )(see Fig. (\ref{3ratio}), and one very short edge denoted $e_{\varepsilon}$ with length $\varepsilon$, the robot is placed by the adversary  adjacent to the middle one of the long edges of the  obstacle.
    So in this case $L_\mathrm{opt}=\frac{\ell}{2}+\varepsilon$. An online algorithm may choose to go either left or right, we assume the adversary will make it harder for any online algorithm and ,therefore, the adversary knows how to place the triangle-obstacle in order to challenge the online, so that the online (any online algorithm) makes an approach to the opposite direction of what is best for it (it is oblivious). Thus, producing a path of length: $\ell+\frac{\ell}{2}$, i.e,
    $$\frac{L_{\on}}{L_\mathrm{opt}}~\leq~\frac{\ell+\frac{\ell}{2}}{\frac{\ell}{2}+\varepsilon}~=~3-\tilde{\varepsilon}$$
\end{proof}
\begin{figure}
  \centering
\newrgbcolor{zzttqq}{0.6 0.2 0.}
\newrgbcolor{qqccqq}{0. 0.8 0.}
\psset{xunit=1.0cm,yunit=1.0cm,algebraic=true,dimen=middle,dotstyle=o,dotsize=3pt 0,linewidth=0.8pt,arrowsize=3pt 2,arrowinset=0.25}
\begin{pspicture*}(3.65,4.3)(10.3,6.7)
\pspolygon[linecolor=zzttqq,hatchcolor=zzttqq,fillstyle=dots*,hatchangle=45.0,hatchsep=0.1](3.92,5.46)(9.54,5.46)(9.54,5.85)
\psline[linecolor=zzttqq](3.92,5.46)(9.54,5.46)
\psline[linecolor=zzttqq](9.54,5.46)(9.54,5.85)
\psline[linecolor=zzttqq](9.54,5.85)(3.92,5.46)
\psline[linewidth=1.6pt](6.76,5.)(9.54,5.46)
\psline[linewidth=1.6pt](9.54,5.46)(9.54,5.85)
\psline[linewidth=1.6pt,linestyle=dashed,dash=1pt 1pt,linecolor=qqccqq](6.76,5.)(3.92,5.46)
\psline[linewidth=1.6pt,linestyle=dashed,dash=1pt 1pt,linecolor=qqccqq](3.92,5.46)(9.54,5.85)
\begin{scriptsize}
\psdots[dotstyle=*,linecolor=blue](9.54,5.85)
\rput[bl](9.62,5.96){\blue{$C$}}
\rput[bl](6.72,5.16){\zzttqq{$l_1$}}
\rput[bl](9.84,5.56){\zzttqq{$\varepsilon$}}
\rput[bl](7.1,6.16){\zzttqq{$l_2$}}
\psdots[dotstyle=*,linecolor=blue](6.76,5.)
\rput[bl](6.42,4.58){\blue{$Robot$}}
\rput[bl](8.42,4.74){$OPT$}
\rput[bl](5.24,5.88){\qqccqq{$Online$}}
\end{scriptsize}
\end{pspicture*}
  \caption{An illustration of a triangular obstacle with an optimal path and online path.}
  \label{3ratio}
\end{figure}
We describe a competitive online algorithm for touring the outside of an open (i.e., excluding the boundary) convex polygon, $\Omega$.

The proposed algorithm suggests that the watchman uses a spiral search (See Fig. \ref{fig:4}.
\begin{figure}
\hspace*{2cm}
\includegraphics[width=0.7\linewidth]{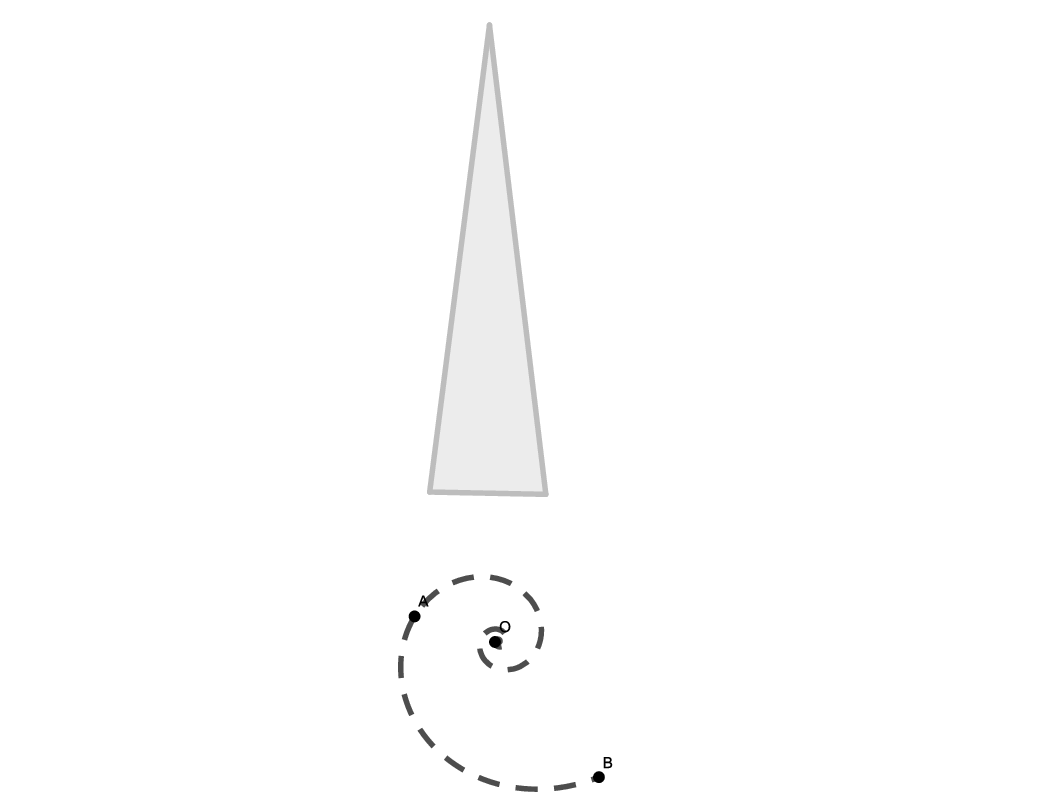}
\caption{ A spiral search which starts at the origin for the same obstacle and leads to a watchman path.\label{fig:1}}
\vskip -0.2in
\end{figure}
 Since the watchman cannot intersect the obstacle, in a case of intersection between the obstacle and the spiral search, we need to define a path which continues the exploration. Assume w.l.o.g that the watchman is located at the origin and that the obstacle is contained in the upper half plane (if it is not, the coordinates can be rotated such that it is).
\begin{definition}
Given a domain $\Omega$ which intersects the spiral search at a point $\left(\tta,r(\tta)\right)$.
Let $A$ be the set of \textbf{entry points}, i.e., the set of points, $\left(\tta,r(\tta)\right)$, such that the spiral search intersects the boundary of the convex domain at a point $\left(\tta,r(\tta)\right)$ and $\left(\tta-\delta,r(\tta-\delta)\right)$ ($\delta \to 0^+$) are not contained in the convex domain, denote these entry points by $\{x_i\}_{i=1}^n$, $x_i\in A$. In a similar way define  $B$ as the set of \textbf{exit points}, i.e. the set of points, $\left(\tta,r(\tta)\right)$, such that  $\left(\tta,r(\tta)\right)$ is not contained in $\Omega$ and $\left(\tta-\delta,r(\tta-\delta)\right)$, $\delta\to 0^+$ are contained in $\Omega$. Denote these exit points by $\{y_i\}_{i=1}^n$
\end{definition}

\begin{notation}
\begin{itemize}
\item Let $l_{\partial\Omega}(x_i,y_i)$ be the length of the shortest path avoiding the convex polygon going clockwise along its boundary from the point $x_i$ to the point $y_i$.
\item Let $C_i$ be the  disk centered at the origin $O$ with radius $r_i$. $\tilde{C}_i$ the upper half the disk.
\item $\gamma(x,y)$ is the path (curve) along the spiral which start at a point $x$ and end at a point $y$.
\end{itemize}
\end{notation}


\begin{algorithm}
\caption{\label{algo}
Watchman path outside convex polygon}
\begin{enumerate}
\item Rotate axes such that $\Omega $ in the upper half plane
\item $E=\left\{\text{all edges  seen from } O \right\}$
\item Move along the logarithmic spiral, $r=e^{b\theta}$ until a new edge  $e$ is found.
\item $E=E\cup \{e\}$
\item If $E$ is a closed polygon then stop
\item If the polygon is hit then move clockwise around the boundary until a new point of the spiral is reached, then continue to step (3)
\item Continue until a watchman path was completed (i.e., when all edges observed form a closed polygon).
\end{enumerate}
\end{algorithm}
\begin{figure}[htp]
\centering
	\includegraphics[width=1.\textwidth]{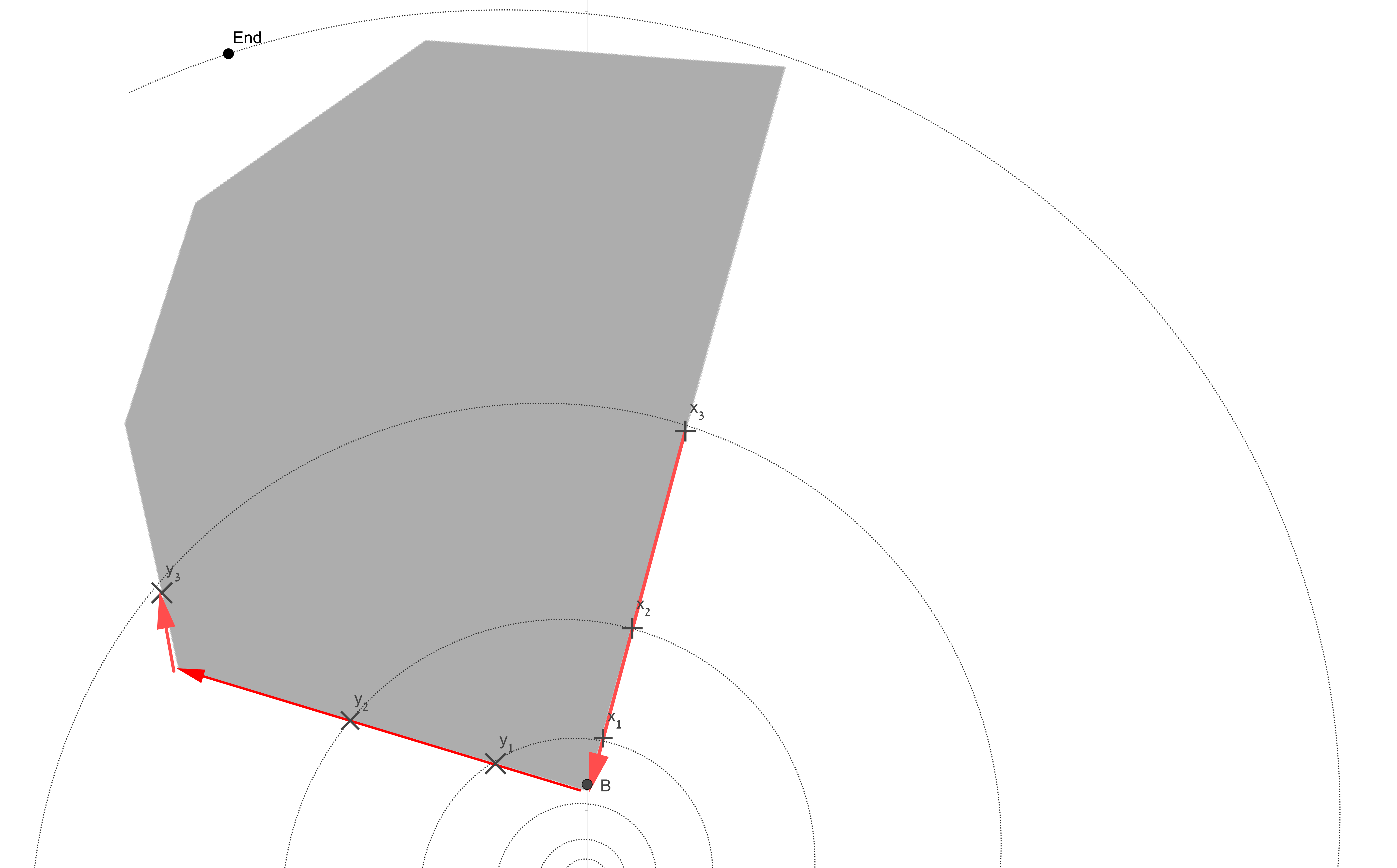}
	\caption{The path suggested by the algorithm. When the spiral search intersects the domain at an entry point, it will move along the boundary of the domain clockwise until it reaches the appropriate exit point of the spiral. Notice that some segments will be traversed more then once. }
\label{fig:4}
\end{figure}
\begin{lemma}
\label{spiral_int}
Given a convex domain $\Omega$ and a spiral search which begins at a given point $O$. If the spiral search intersects the boundary of the domain at a point $x_i \in A$, and let $y_i\in B$ s.t $y_i:=(\theta_i,R_i)$ be the appropriate exit point, the arc length $l_{\partial\Omega}(x_i,y_i)\leq (\pi+2)\cdot R_i$.
\end{lemma}
\begin{figure}[htp]
\centering
\includegraphics[width=0.7\textwidth]{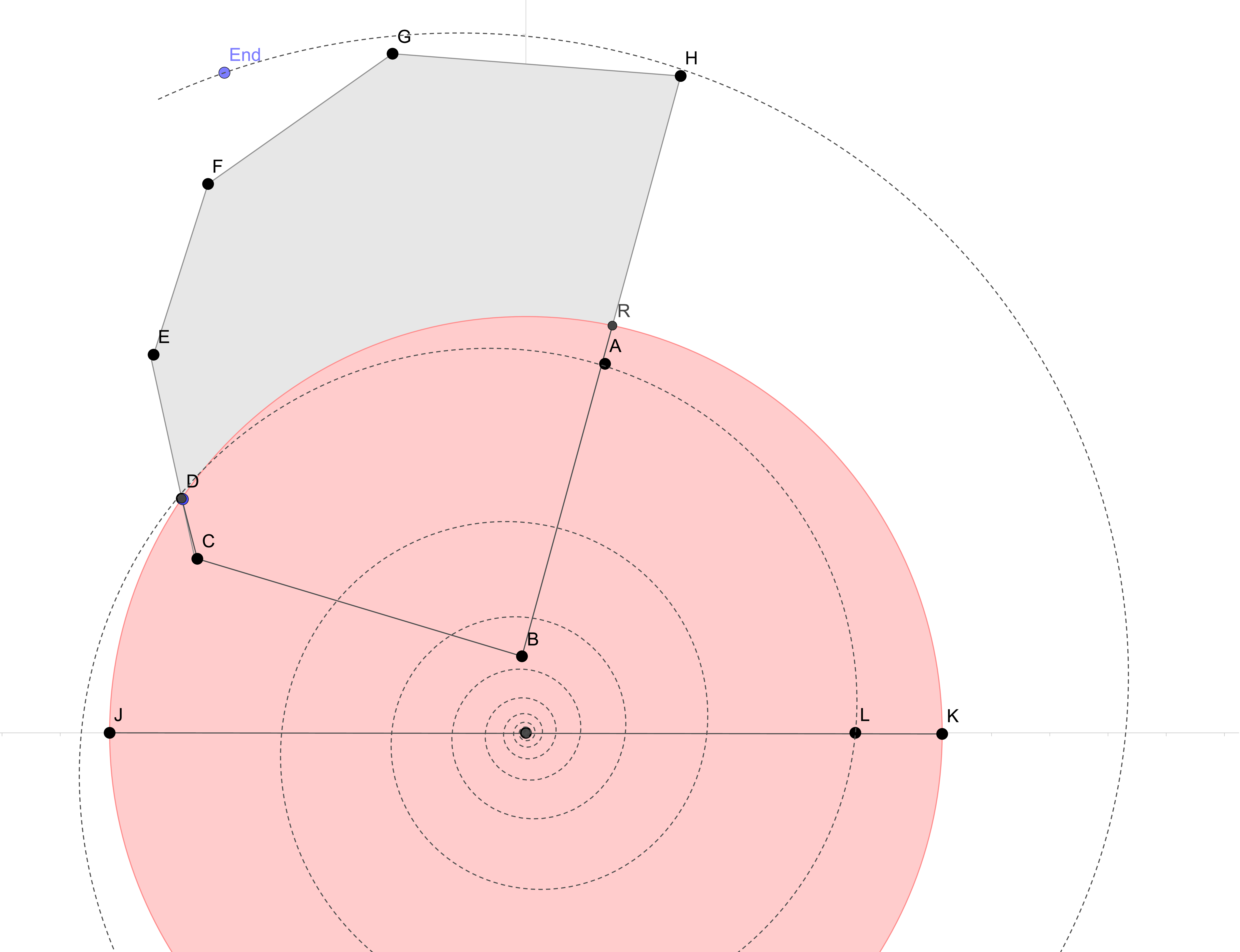}
	\caption{Bounding the length of the path along the polygon by the circumference of a semicircle with radius $R$, defined by the intersection of the convex domain and the spiral.}
\label{fig:fig6}
\end{figure}
\begin{proof}
We will define a convex domain whose boundary is
\begin{align*}
P_i:=\gamma(x_i,y_i)\cup \partial\Omega(x_i,y_i) \ ,\quad 1 \leq i \leq n \;,
\end{align*}
where $\gamma(x_i,y_i)$ is the curve along the spiral which starts at the point $x_i$ and ends at the point $y_i$ (with the respective length of the spiral arc being $l_s(x_i,y_i)$) and $\partial\Omega(x_i,y_i)$ is the curve going clockwise along the polygon which start at the point $x_i$ and ends at the point $y_i$ (with the respective length $l_{\partial\Omega}(x_i,y_i)$), see Fig. \ref{fig:fig6}.
Since $d(y_i,O)=R_i$ (the Euclidean distance  between the point $(\theta,R_i)$ and the origin), the domain $P_i$ is contained in the half  $\tilde{C_i}$ with perimeter $\frac{2 \pi R_i}{2}+2\cdot R_i=(\pi+2)\cdot R_i$.

Since the domains $P_i$ and $\tilde{C_i}$ are convex, by Lemma \ref{boundary length}, then
$ |\partial{P_i} | \leq |\partial \tilde{C_i}|= (\pi+2)R_i$.

So we get that $l_{\partial\Omega}(x_i,y_i) \leq |(P_i)| \leq (\pi+2)R_i$.
\end{proof}

\begin{notation}
Denote the length of the curve obtained by the algorithm which ends at a point $y=( \theta,R)$ by $L_\mathrm{alg}(R)$.
\end{notation}

\begin{lemma}
The  ratio between the length of the path constructed by Algorithm~\ref{algo} and the spiral path is given by
\label{lemma}
 $$ L_\mathrm{alg}(R)\leq \left(1+ \frac{(2+\pi)}{(e^{2 \pi}-1) \sqrt{1+\frac{1}{b^2}}}\right) \cdot L_s(R)\;.$$
\end{lemma}

\begin{proof}
The algorithm starts a spiral search until it  intersects the boundary of the convex obstacle. The spiral intersects the convex domain during the $i$th lap when the angle is at most $\tta\le 2(i-1)\pi$  for some $i\in\mathbb{Z}$. The respective distance from the origin is $e^{(2i-1)b\pi}<R$ for every $-\infty <i\leq N$. By Lemma (\ref{spiral_int}) for every intersection $x_i\in A \wedge y_i \in B$ ($ i \leq N$) with a distance $R_i$ :
$$l_{\partial\Omega}(x_i,y_i)|< | P_i| \leq (\pi+2) \cdot R_i$$
and the  ratio is (assuming $b \geq 0$)
\begin{eqnarray*}
 &L_\mathrm{alg}& < L_s(R) + \sum\limits_{i=-\infty}^N (2+\pi) \cdot R_i = L(R)+\sum\limits_{i=-\infty}^N (2+\pi) \cdot e^{(2i-1)b\pi}\\
 &=&L_s(R)+(2+\pi)e^{b\pi}\cdot  \frac{e^{2 \pi bN}}{e^{2 \pi b}-1} \leq \left( 1+ \frac{(2+\pi)}{(e^{2 \pi}-1) \sqrt{1+\frac{1}{b^2}}}\right)\cdot L_s(R) \ .
\end{eqnarray*}
\end{proof}

Now, we are ready to find the desired competitive ratio between $L_\mathrm{alg}$  and $L_\mathrm{opt}$  (our algorithm vs the optimal).

In order to compare the algorithm's performance to the optimal off-line algorithm, we notice that every watchman path
necessarily visits (at least one point of) each half-plane in $\mathbb{R}^2\setminus \Omega$ that is formed by an edge of $\Omega$.
Thus, the length of the optimal offline watchman path is at least $h=\max_i |h_i|$ where $h_i$ is the shortest path to the half plane defined by edge $i$
of $\Omega$ that does not intersect $\Omega$ (i.e. lies outside $\Omega$).
We use $\gamma_\mathrm{alg}(p_1,p_2)$ to denote the path of the algorithm between the points $p_1$ and $p_2$.

\begin{lemma}
\label{main_lem}
Let $H_i$ be the external half plane defined by the $i$th edge of $\Omega$, such that $O\notin H_i$, and let $h_i$ be the
shortest path between $O$ and $H_i$ in $\mathbb{R}^2\setminus \Omega$. Let
\begin{align}
\label{eq:points}
N=\left\lceil\frac{\ln |h_i|}{2\pi b}\right\rceil \ ,
\end{align}
Then $\gamma_{alg}\left(\left(\tta=2\pi N,r=e^{b\tta}\right),\left(\tta=2\pi N+\pi,r=e^{b\tta}\right)\right)$ visits  $H_i$.
\end{lemma}
\begin{proof}
Let $p_i$ be the point in $H_i$ which is closest to $O$ in $\mathbb{R}^2\setminus \Omega$.

Now, assume that the shortest path
to $p_i$ goes counter clockwise relative to $\Omega$. Let $S$ be the infinite ray from $O$ towards $p_i$. The spiral
$\gamma\left((\tta=2\pi N,r=e^{b\tta}),(\tta'=2\pi N+\pi,r=e^{b\tta'})\right)$ intersects $S$
 at some point $p$. We claim that the path $\gamma((\tta=2\pi N,r=e^{b\tta}),p)$ does not
intersect $\Omega$. Indeed, if $\Omega$ is entirely contained in the cone with angles between the angle of $S$ and
$\pi$ then the spiral segment $\gamma\left((\tta=2\pi N,r=e^{b\tta}),(\tta'=2\pi N+\pi,r=e^{b\tta'})\right)$
cannot intersect $\Omega$. Otherwise, assume that $\gamma\left((\tta=2\pi N,r=e^{b\tta}),(\tta'=2\pi N+\pi,r=e^{b\tta'})\right)$
intersects $\Omega$. Let $q$ be the point of intersection. Let $C$ be the convex region whose boundary is
$h_i\cup \overline{O p_i}$, and let $s\in \Omega\cap C$ be a point in the intersection. The segment $\overline{qs}$
has both endpoints in $\Omega$, but intersects $S$, in contradiction to convexity.
Since $O\notin H_i$ and since $H_i$ is convex, it follows that $p$ is in $S\cap H_i$ and thus in $H_i$.

Now suppose the shortest path
to $p_i$ goes  clockwise relative to $\Omega$. let $S$ be the infinite ray from $O$ towards $p_i$. the spiral
$\gamma\left((\tta=2\pi N,r=e^{b\tta}),(\tta'=2\pi N+\pi,r=e^{b\tta'})\right)$ intersects $S$
 at some point $p$. We claim that $p$ lies outside $\Omega$ and thus also lies on
$L_\mathrm{alg}\left((\tta=2\pi N,r=e^{b\tta}),(\tta'=2\pi N+\pi,r=e^{b\tta'})\right)$.
Indeed, if $\Omega$ is fully contained in the cone between the angle 0 and the angle of $S$,
then clearly the ray $S$ is in free space. If $S$ intersects $\Omega$, then let $q\in S\cap \Omega$
be a point then $q\in\Omega$. Assume that $p \in \Omega$ then $p_i\in \overline{pq}$, but $p_i\notin\Omega$
in contradiction to convexity.
Again, we have $p\in H_i$.
\end{proof}

Thus we have,
\begin{theorem}
Let
\begin{equation}
\label{eq:1}
N=\left\lceil\frac{\ln h}{2\pi b}\right\rceil \ .
\end{equation}
The path of the algorithm which stats at a point $O$ and end at a point $\left(\tta=2\pi N+\pi,r=e^{b\tta}\right)$
visits all half planes defined by edges of $\Omega$, and thus is a watchman path.
\end{theorem}
\begin{proof}
By Lemma \ref{main_lem} the path visits each of the half planes at an appropriate radius, and thus is a watchman path.
\end{proof}

\begin{theorem}
The algorithm has a competitive ratio of $24.35\ldots$ compared to the optimal offline watchman path.
\end{theorem}
\begin{proof}
Since a watchman path must visit every half plane, the length of the optimal watchman path
is at least $h$. By (\ref{eq:1}) and Lemma (\ref{lemma}), taking another lap of the spiral ensure thats watchman path is obtained. Thus
$$
L_\mathrm{alg}\left(e^{(2\pi N+\pi)b}\right)\leq C \cdot    L_s\left(e^{(2\pi N+\pi)b}\right)\ ,
$$
where $C=1+ \frac{(2+\pi)}{(e^{2 \pi}-1) \sqrt{1+\frac{1}{b^2}}}$.
By (\ref{eq:1}), we have
$$
N=\left\lceil\frac{\ln h}{2\pi b}\right\rceil \leq\frac{\ln h+1}{2\pi b}
$$
If $L_\mathrm{opt}(R)$ is the optimal path then
\begin{eqnarray}
 L_\mathrm{alg}(R) & \leq & L_\mathrm{alg}\left(e^{(2\pi N+\pi)b}\right) \nonumber\\
&\leq& \left( 1+ \frac{(2+\pi)}{(e^{2 \pi}-1)\cdot \sqrt{1+\frac{1}{b^2}}}\right) \cdot L_s\left(e^{(2\pi N+\pi)b}\right) \nonumber\\
&\leq&\left( 1+ \frac{(2+\pi)}{(e^{2 \pi}-1) \sqrt{1+\frac{1}{b^2}}}\right)  \sqrt{1+\frac{1}{b^2}}\cdot e^{{(1+\ln h)}+\pi b} \nonumber\\
	&=&  \left(\sqrt{1+\frac{1}{b^2}}+ \frac{2 +\pi}{e^{2 \pi}-1}\right)e^{\pi b+1}\cdot h \nonumber\\
	&\leq& \left(\sqrt{1+\frac{1}{b^2}}+ \frac{2 +   \pi}{e^{2 \pi}-1}\right)e^{\pi b+1} \cdot L_\mathrm{opt}
  \label{min}
\end{eqnarray}
minimization by $b$ leads to $b=0.29$ and  $L_\mathrm{alg}\lesssim 24.3504 \cdot L_\mathrm{opt}$.
\end{proof}

\section{Further Results}
\begin{figure}[htp]
\vskip -.2in
\centering
\includegraphics[width=0.75\textwidth]{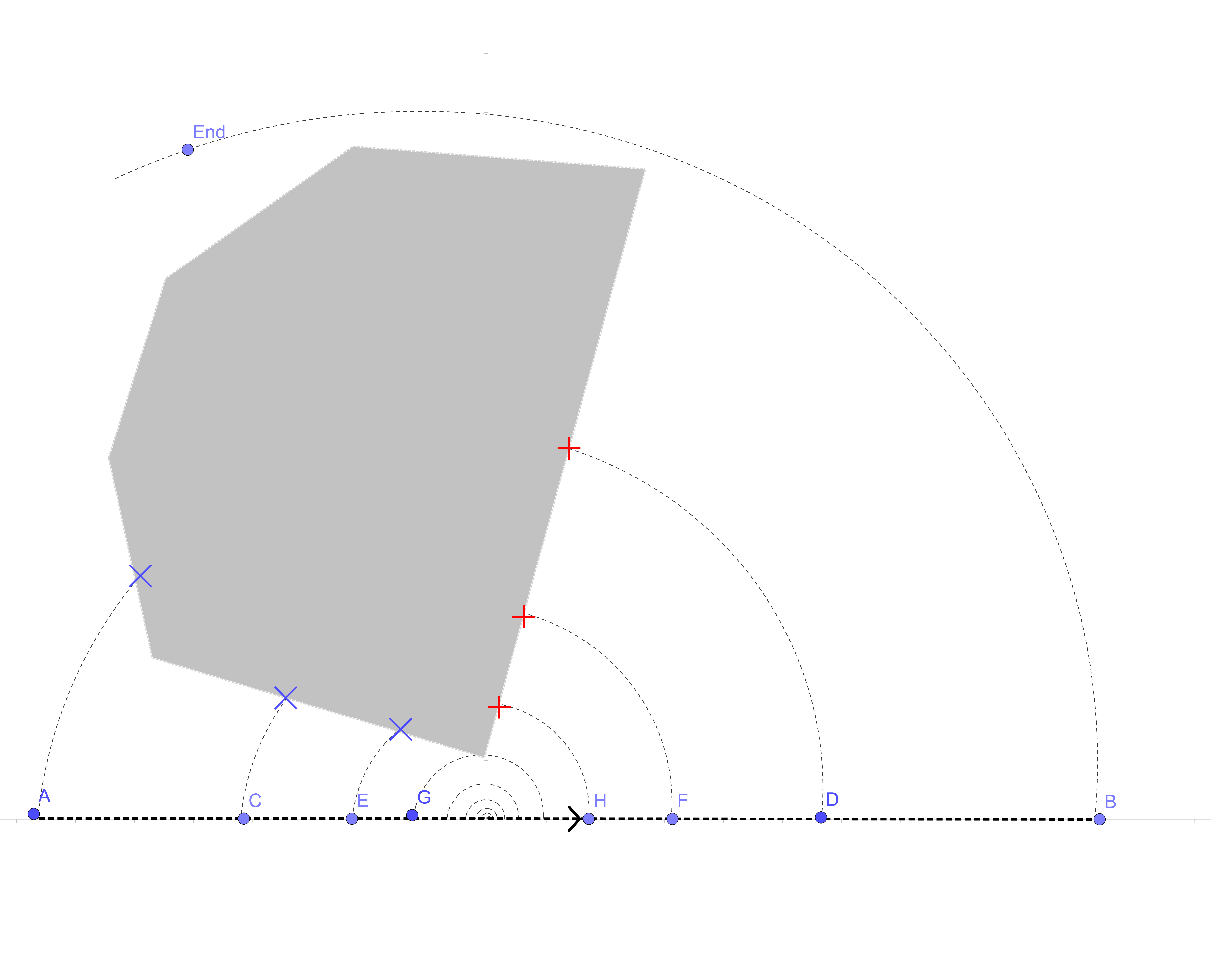}
	\caption{The path generated by the second proposed algorithm. In this case in every lap the lower half of the spiral is replaced by line segments, with the respective segments in this case being $\left\{AB, SD, EF, GH,\dots \right\}$ etc. }
\label{fig:top and tailed}
\end{figure}

Since spiral search encircles the origin, the last algorithm can  be improved by touring only along the parts of the spiral that have $0<\tta \mod 2\pi<\pi$ and replacing the bottom parts of the spiral by straight line segments (see Fig.\ref{fig:top and tailed}). The arc length of the spiral can be computed by the formula
\begin{align}
L_s(R)=\int_{R_0}^{R_1} \sqrt{1+r^2\left(\frac{d\theta}{dr}\right)^2}dr{=}\sqrt{1+\frac{1}{b^2}}\left(R_1-R_0\right)
\end{align}
The arcs between the points $(\pi+2\pi i ,R)$ and $(2 \pi +2 \pi i,R)$ where $ i \leq N$, which will be replaced by the length $\left|(O,R(\pi+2\pi i)\right|+\left|(O,R(2 \pi+2\pi i)\right|$ (where $O$ is the origin). So the length of this semi-spiral is
\begin{multline}
\label{eq:3}
\sum_{i=-\infty}^N  \left(\sqrt{1+\frac{1}{b^2}}\left( e^{b(2 \pi+2  \pi i)}-e^{b( \pi+2 \pi i)}\right)-\left( e^{b( \pi+2 \pi i)}+ e^{b(2 \pi+2\pi i)}\right)\right)= \\
\left(\sqrt{1+\frac{1}{b^2}} \left( e^{2 \pi b}-e^{\pi b} \right)-\left(  e^{2 \pi b}+ e^{\pi b}\right)\right)\cdot\left(\frac{ e^{2 \pi b N}}{e^{2 \pi b}-1}\right) \, .
\end{multline}

This progression by a semi-spiral reduces the length of the path produced by the new algorithm. This length will be denoted by $L_{\mathrm{alg}_2}$.
\begin{theorem}
$L_{\mathrm{alg}_2}$ gives a $22.77\ldots$ competitive ratio respective to the optimal algorithm
\end{theorem}
\begin{proof}
\begin{eqnarray}
\label{eq:2}
L_{\mathrm{alg}_2}(R)&\leq& \left(\sqrt{1+\frac{1}{b^2}}+ \frac{2 +   \pi}{e^{2 \pi}-1}\right)e^{\pi b+1}\cdot h- \left({\sqrt{1+\frac{1}{b^2}}} \left( e^{2 \pi b}-e^{\pi b} \right)-\left(  e^{2 \pi b}+ e^{\pi b}\right)\right)\cdot\left(\frac{ e^{2 \pi b N}}{1-e^{-2 \pi b}}\right)\nonumber\\
&=& \left(\sqrt{1+\frac{1}{b^2}}+ \frac{2 +   \pi}{e^{2 \pi}-1}\right)e^{\pi b+1}\cdot h- \left( \left( e^{2 \pi b}-e^{\pi b} \right)-  \frac{e^{2 \pi b}+ e^{\pi b}}{{\sqrt{1+\frac{1}{b^2}}}}\right)\cdot{\sqrt{1+\frac{1}{b^2}}}\left(\frac{ e^{2 \pi b N}}{1-e^{-2 \pi b}}\right) \nonumber\\
&=&\left(\sqrt{1+\frac{1}{b^2}}+ \frac{2 +   \pi}{e^{2 \pi}-1}\right)e^{\pi b+1}\cdot h- \left( \left( e^{2 \pi b}-e^{\pi b} \right)-  \frac{e^{2 \pi b}+ e^{\pi b}}{{\sqrt{1+\frac{1}{b^2}}}}\right)\cdot\left(\frac{ \gamma(2 \pi N)}{1-e^{-2 \pi b}}\right) \nonumber\\
&\leq&\left(\sqrt{1+\frac{1}{b^2}}+ \frac{2 +   \pi}{e^{2 \pi}-1}\right)e^{\pi b+1}\cdot h- \left( \left( e^{2 \pi b}-e^{\pi b} \right)-  \frac{e^{2 \pi b}+ e^{\pi b}}{{\sqrt{1+\frac{1}{b^2}}}}\right)\cdot\left(\frac{ h}{1-e^{-2 \pi b}}\right)\nonumber\\
&\leq&\left(\left(\sqrt{1+\frac{1}{b^2}}+ \frac{2 +   \pi}{e^{2 \pi}-1}\right)e^{\pi b+1} - \left( \left( e^{2 \pi b}-e^{\pi b} \right)-  \frac{e^{2 \pi b}+ e^{\pi b}}{{\sqrt{1+\frac{1}{b^2}}}}\right)\cdot\left(\frac{ 1}{1-e^{-2 \pi b}}\right)\right)\cdot L_\mathrm{opt}
\end{eqnarray}
a minimum is obtained at the point $b=0.2929150042$ giving $L_{\mathrm{alg}_2}(R)\leq 22.7749 \cdot L_\mathrm{opt}$.
\end{proof}
\bibliographystyle{spmpsci}
\bibliography{Thesis1,shai_bib}

\end{document}